# Ce$^{3+}$:CaSc$_2$O$_4$ Crystal Fibers for Green Light Emission: Growth Issues and Characterization

Detlef Klimm, Jan Philippen, Toni Markurt, and Albert Kwasniewski

Leibniz Institute for Crystal Growth, Max-Born-Str. 2, 12489 Berlin, Germany

## ABSTRACT

Ce$^{3+}$ is known to show broad optical emission peaking in the green spectral range. For the stabilization of 3-valent cerium in ceramic phosphors such as calcium scandate CaSc$_2$O$_4$, often co-doping with sodium for charge compensation is performed (Na$^+$, Ce$^{3+}$ ↔ 2 Ca$^{2+}$). At the melting point of CaSc$_2$O$_4$ (≈2110°C), however, alkaline oxides evaporate completely and co-doping is thus no option for crystal growth from the melt. It is shown that even without co-doping Ce$^{3+}$:CaSc$_2$O$_4$ crystal fibers can be grown from the melt by laser-heated pedestal growth (LHPG) in a suitable reactive atmosphere. Reactive means here that the oxygen partial pressure is a function of temperature and $p_{O2}(T)$ rises for this atmosphere in such a way that Ce$^{3+}$ is kept stable for all $T$. Crystal fibers with ≈1 mm diameter and ≤50 mm length were grown and characterized. Differential thermal analysis (DTA) was performed in the pseudo-binary system CaO–Sc$_2$O$_3$, and the specific heat capacity $c_p(T)$ of CaSc$_2$O$_4$ was measured up to 1240 K by differential scanning calorimetry (DSC). Near and beyond the melting point of calcium scandate significant evaporation of calcium tends to shift the melt composition towards the Sc$_2$O$_3$ side. Measurements and thermodynamic calculations reveal quantitative data on the fugacities of evaporating species.

## INTRODUCTION

The direct conversion of electric energy to visible light is one of the big research challenges nowadays. From the 1960s on, Ga(As,P) based light emitting diodes (LED's) became available which were used mainly as indicators with red emission. Not before 1994, (Ga,In)N based LED's were developed for the blue and even ultraviolet spectral range [1]. Unfortunately, just the green spectral range where human eyes have maximum sensitivity is hard to reach by direct emission, because material systems mentioned above suffer from reduced electro-optical conversion efficiency in the center of the visible spectral range. Irrespective of some recent progress with (In,Ga)N layers on GaN substrates [2], still frequency conversion from UV to the green spectral region by photoluminescence is the most important process for the generation of green light, especially in white LED's for general illumination.

Ce$^{3+}$ ions in oxide environment can convert UV light efficiently to the green, with maximum intensity around 515 nm [3], and calcium scandate CaSc$_2$O$_4$ can be doped with several trivalent rare earth ions, among them Ce$^{3+}$ [4,5]. The octahedral radii (Shannon) of Ce$^{3+}$ (115 pm) and Ca$^{2+}$ (114 pm) are almost identical, and it can be assumed that Ce$^{3+}$ substitutes for Ca$^{2+}$ rather than for the much smaller (88.5 pm) Sc$^{3+}$. For the production of white LED's, ceramic phosphors are used. The incorporation of Ce$^{3+}$ into the CaSc$_2$O$_4$ matrix is then enabled by co-doping with the similarly sized Na$^+$ (octahedral radius 116 pm), and Na$^+$ + Ce$^{3+}$ substitute for 2 Ca$^{2+}$ conserving charge neutrality.

In a recent study we reported the growth of $Ce^{3+}$:$CaSc_2O_4$ single crystals from the melt [6]. At the very high fusion temperature $T_f = 2110°C$ of this material the fugacity of $Na_2O$ exceeds ambient pressure and the volatility of sodium is so high that it evaporates from the melt completely. It will be shown here that a pure (99.999%) nitrogen atmosphere delivers a suitable background oxygen partial pressure that maintains $Ce^{3+}$ under growth conditions without co-doping.

**EXPERIMENT AND DISCUSSION**

The binary phase diagram $CaO–Sc_2O_3$ is not yet published and ca. 20 compositions ranging from pure $Sc_2O_3$ to pure CaO were studied by differential thermal analysis (DTA) with a NETZSCH STA 429 up to a maximum temperature (dependent on composition) of 2200°C. Even if this equipment is capable for measurements up to 2400°C, calcium evaporation became severe if the samples were heated significantly beyond the fusion point of single phase $CaSc_2O_4$ which was found at 2110°C, in total agreement with the literature [7]. Left and right from $CaSc_2O_4$, two eutectics were found at 1965±5°C ($Sc_2O_3/CaSc_2O_4$) or 1953±5°C ($CaSc_2O_4$/CaO), respectively.

Besides, XRD measurements of $(CaO)_x(Sc_2O_3)_{1-x}$ powder mixtures with $0 \leq x \leq 1$ were performed after annealing them at 1600°C for 48 h. All peaks of the spectra could be assigned to the phases $Sc_2O_3$, $CaSc_2O_4$, CaO, or $Ca(OH)_2$, and the latter is obviously a secondary product formed from CaO with traces of water that are ubiquitous also during X-ray measurements (Fig. 1). Neither longer annealing times, nor XRD measurements with powdered single crystalline fibers revealed other phases.

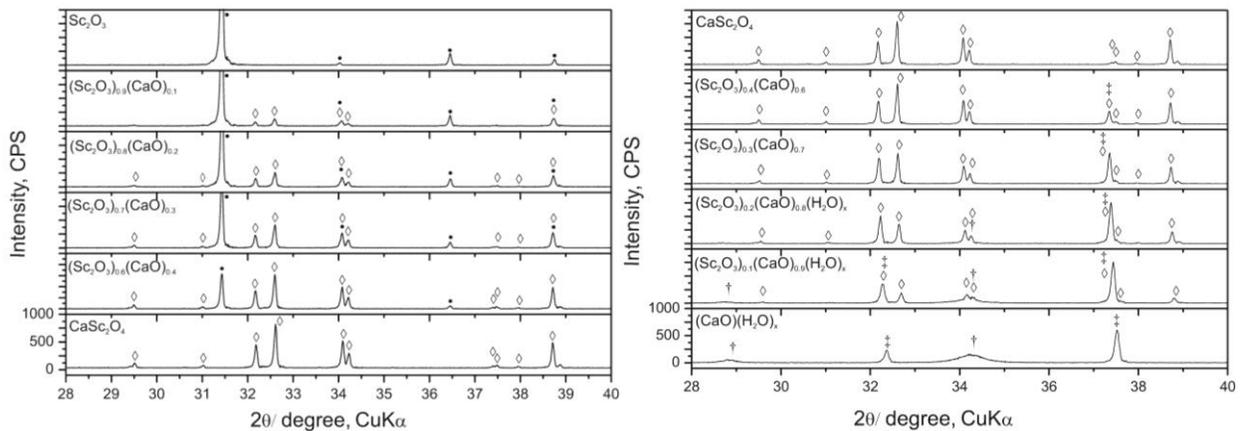

**Fig. 1: Left: XRD patterns of sintered compositions within the range of pure scandium oxide (top) and calcium scandate (bottom). Reflections referring to $Sc_2O_3$ are marked with •, $CaSc_2O_4$ reflections are marked with ◊. Right: XRD patterns for the composition range from $CaSc_2O_4$ (top) to CaO (bottom). CaO reflections = ‡, $Ca(OH)_2$ = †.**

It should be noted that no variation of the $Sc_2O_3$ lattice parameters with $x$ could be detected, whereas [7] reported 3.1% CaO solubility in $Sc_2O_3$. For CaO and $CaSc_2O_4$ the formation of solid solutions with a small (ca. 3%) homogeneity width seems to be probable. These experimental results allow proposing the phase diagram that is shown in Fig. 2. There $ScO_{1.5}$ rather than $Sc_2O_3$ was used as component to assure that the cation number on the concentration axis is homogeneous. Then the liquidus slope on both sides of the intermediate phase is the same and it turns out that it is steeper compared to that of the components. This results from the comparably small heat of fusion $\Delta H_f$ for $CaSc_2O_4$: Exact data are not available, but comparing the fusion peak area of $CaSc_2O_4$ with that of $Al_2O_3$ ($T_f$ = 2054°C, only 56 K lower), allowed to estimate $\Delta H_f \approx 196$ kJ/mol for $CaSc_2O_4$, which corresponds to 65 kJ per mol of one cation. For comparison, $\Delta H_f$ of CaO is much larger (79.5 kJ/mol). The specific heat capacity of $CaSc_2O_4$ was measured by DSC and is $c_p = 140.5441 + 0.0272\,T - 1.462047 \times 10^{-6}\,T^2$ ($T$ in K, [10]).

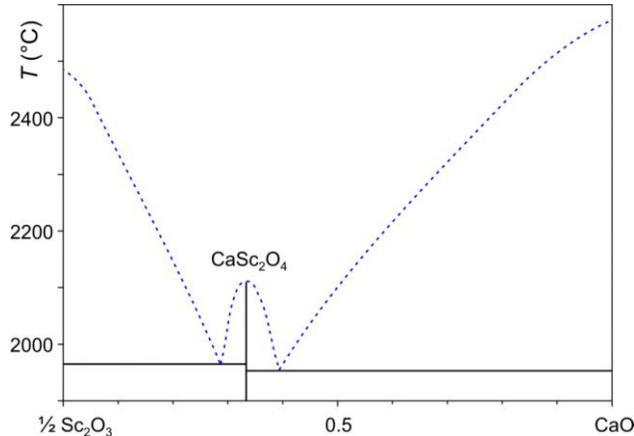

**Fig. 2: Tentative phase diagram CaO–$Sc_2O_3$ with the single intermediate phase $CaSc_2O_4$.**

Undoped and cerium doped (1% Ce substitutes for Ca) $CaSc_2O_4$ powder mixtures were prepared by mixing of $CaCO_3$, $Sc_2O_3$, and $CeO_2$ powders in stoichiometric ratio. Starting materials were of 99.99% purity, and a first carbonate calcination run was performed at 900°C for 8 hours. Subsequent grinding and annealing at 1200°C in air was performed until no other phases except $CaSc_2O_4$ were found by X-ray powder diffraction. In the doped powders additionally $CeO_2$ peaks could be detected, irrespective of the low doping concentration. Obviously $Ce^{4+}$ remains stable under the annealing conditions, and the large scattering cross section of the heavy cerium ion was sufficient for detection by XRD. The reacted powders were cold isostatically pressed at 2 kbar, annealed 24 h at 1600°C and cut into rectangular prisms which were used as feed rods for the LHPG growth.

The LHPG (laser-heated pedestal growth) technique was developed in the 1980's [8,9]. For the growth, the top of the vertical ceramic feed rod was heated by focused laser light. A seed crystal from the top moved downwards until it touches the molten top of the feed. Subsequently, both feed and seed are moved almost simultaneously upwards and the growth process is performed. Details can be found elsewhere [10]. It should be noted that LHPG crystal growth is performed without direct contact to a container. This is beneficial because nearly arbitrary atmospheric conditions can be used. The severe influence of atmosphere could be seen here directly during growth (Fig. 3): If the growth was performed in a slightly oxidizing atmosphere (Fig. 3 middle), the melt zone ($T_f$=2110°C) glows almost white. After growth, $CeO_2$ can be identified in the $CaSc_2O_4$ matrix of the fiber by XRD. Growth in pure nitrogen (99.999% $N_2$) or under even stronger reducing conditions (up to 5% $H_2$ in $N_2$ were used) leads to a molten zone that glows red (left hand side). No cerium oxide(s) could be found in such fibers (see (Fig. 3 right) by XRD, which is an indication that indeed $Ce^{3+}$ can be incorporated into the $CaSc_2O_4$ crystal structure. TEM-EDX analysis indicates that $Ce^{3+}$ sits on $Ca^{2+}$ lattice sites [10]. In contrast, $Ce^{4+}$ which is formed under oxidizing conditions cannot substitute $Ca^{2+}$.

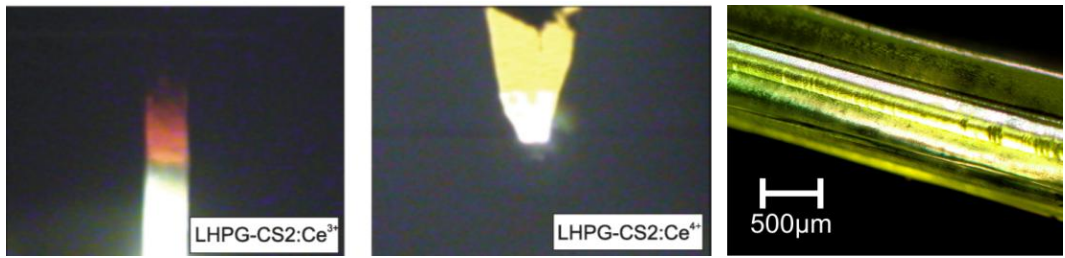

**Fig. 3: LHPG process with growing CaSc$_2$O$_4$ fibers (1% cerium doping, fiber diameter ca. 1 mm). The left photograph was taken during growth in nitrogen (99.999% purity), and the middle photograph was taken in oxidizing atmosphere (2% O$_2$ in Ar). The Ce$^{3+}$ doped fiber (after cooling to room temperature) in the right photograph [6] is reprinted with permission from Elsevier.**

Photoluminescence spectra of Ce:CaSc$_2$O$_4$ single crystal LHPG fibers were already reported [6] and show with 337 nm excitation luminescence peaking at 515 nm, which is a clear indication for Ce$^{3+}$. The more reducing the growth atmosphere is, the stronger appears the green emission, and the strongest 515 nm peak was observed with fibers grown in forming gas (95% N$_2$ + 5% H$_2$).

Unfortunately, the volatility of species is independent on atmosphere only, if the valence state is identical in the condensed and gaseous state. Fig. 4 shows this for 3 different atmospheres. All gases were assumed to have 99.999% (5N) purity which was used in our experiments. Air was assumed to be the rest gas impurity, resulting in a background oxygen partial pressure up to $2\times10^{-6}$ bar which cannot be avoided. It should be noted that in nitrogen the oxygen partial pressure is a function of temperature, because nitrous oxides NO$_x$ are formed at low $T$. Only for CaO, which is present as such in all phases the fugacity does not depend on atmosphere. Ca is created from CaO by reduction, and its fugacity becomes larger in the order Ar < N$_2$ < forming gas. In forming gas the Ca fugacity approaches 1 mbar, which impedes crystal growth significantly due to strong evaporation. Besides, calcium(I) hydroxide CaOH contributes to evaporation under such conditions.

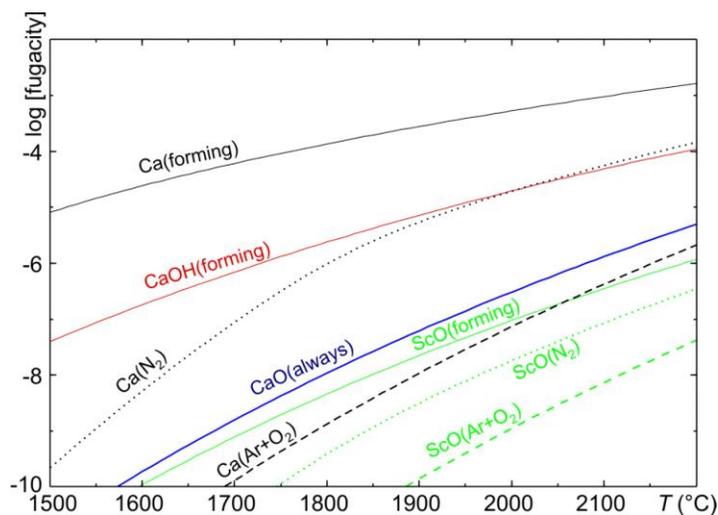

**Fig. 4: Fugacity of main species over CaO+Sc$_2$O$_3$ in different atmosphere from oxidizing (Ar+2% O2) over almost inert (N$_2$) to strongly reducing (forming gas N$_2$ + 5% H$_2$).**

# CONCLUSIONS

Cerium doped $CaSc_2O_4$ fibers can be grown from the melt, and under reducing conditions cerium is built in as $Ce^{3+}$. Co-doping by alkaline elements is not required. Nitrogen atmosphere with 99.999% purity is a compromise where a significant share of cerium occurs as $Ce^{3+}$ (see Fig. 3), and calcium evaporation remains on such a low level that stable LHPG crystal growth can be maintained.

$CaSc_2O_4$ is the only intermediate compound in the $CaO$–$Sc_2O_3$ quasibinary system, which is in significant difference to the more complicated $BaO$–$Sc_2O_3$ system [11]. The congruent melting point of $CaSc_2O_4$ of 2110°C is within the range were Czochralski crystal growth from iridium crucibles is feasible, and hence $Ce^{3+}$:$CaSc_2O_4$ bulk crystals could be a material for future green solid state lasers.

# ACKNOWLEDGMENTS

The authors express their gratitude to C. Guguschev and R. Bertram for collaboration.